\documentclass[12pt]{article}
\usepackage{graphicx}

\def\Title#1{\begin{center} {\Large #1 } \end{center}}

\newenvironment{Abstract}{\begin{quotation}  }{\end{quotation}}
\newenvironment{Presented}{\begin{quotation} \begin{center} 
             PRESENTED AT\end{center}\bigskip 
      \begin{center}\begin{large}}{\end{large}\end{center} \end{quotation}}
\def\Acknowledgements{\bigskip  \bigskip \begin{center} \begin{large}
             \bf ACKNOWLEDGEMENTS \end{large}\end{center}}

\def\beq{\begin{equation}}
\def\eeq#1{\label{#1}\end{equation}}
\def\eeqn{\end{equation}}

\def\beqa{\begin{eqnarray}}
\def\eeqa#1{\label{#1}\end{eqnarray}}
\def\eeqan{\end{eqnarray}}

\let\bar=\overbar

\def\Dslash{\not{\hbox{\kern-4pt $D$}}}
\def\dslash{\not{\hbox{\kern-2pt $\del$}}}

\def\msb{{\bar{\ssstyle M \kern -1pt S}}}

\begin{document}
\begin{titlepage}

\vfill
\Title{Observing dark photon with dark matter detectors}
\vfill

\begin{center}

{\bf \sc Haipeng An$^{\,(a)}$, Maxim Pospelov$^{\,(a,b)}$ and Josef Pradler$^{\,(c)}$}

\smallskip
\medskip

$^{\,(a)}${\it Perimeter Institute for Theoretical Physics, Waterloo,
ON, N2L 2Y5, Canada}

$^{\,(b)}${\it Department of Physics and Astronomy, University of Victoria, \\
  Victoria, BC, V8P 1A1 Canada}

$^{\,(c)}${\it Department of Physics and Astronomy, Johns Hopkins University, \\
  Baltimore, MD, 21210 USA}

\end{center}

\vfill
\begin{Abstract}
Motivated by various kinds of new physics models, a new light neutral vector boson (the dark photon) connected to the Standard Model of particle physics only through the kinetic mixing with the $U(1)_Y$ factor has been studied extensively. Various kinds of experiments are proposed to detect it. For the dark photon with a mass smaller than 10 keV, it can be copiously produced inside the Sun, and then be detected by the detectors on the earth. We show that with the S2 only analysis, the result from XENON10 experiment provides the up-to-date most stringent limit on the region $10^{-5}~{\rm eV} < m_V < 10^{3}$ eV: $\kappa \times m_V < 3\times 10^{-12}$ eV for the St\"uckelberg dark photon model, where $\kappa$ is the kinetic mixing and $m_V$ is the mass of the dark photon. If there is a light Higgs boson accompanied with the dark photon, the sensitivity of XENON10 experiment is $\kappa \times e' < 10^{-13}$, which is still second to the constraints from the lifetime of horizontal branch starts which dictates $\kappa \times e' < 3\times10^{-14}$.

\end{Abstract}
\vfill
\begin{Presented}
CosPA 2013 Symposium\\
Honolulu, Hawai'i, USA, Nov 12-15, 2013
\end{Presented}
\vfill
\end{titlepage}
\def\thefootnote{\fnsymbol{footnote}}
\setcounter{footnote}{0}

\section{Introduction}

The Standard Model of particle physics (SM) can naturally be extended with a light neutral vector particle with only kinetic mixing to the photon field at low energy. This extension has motivations from both the theoretical and phenomenological considerations. Theoretically, most of the string theory constructions have remnants of a light neutral vector field left with all the heavy modes integrated out. Phenomenologically, it is well-motivated to use the dark photon as a portal to connect dark matter to the SM sector. For MeV-scale dark photons, beam-dump experiments are proposed to the detect it directly by shooting electron or proton beams into a fixed target. A recent review on searching for MeV-scale dark photon can be found in Ref.~\cite{Essig:2013lka}. For a Sub-keV dark photon, it can be produce inside the stellar systems, which can influence the chain of nuclear reaction processes inside the center of  the stars, and therefore be constrained by the stellar lifetime. For  a sub-eV scale dark photon, it can also be produced in a terrestrial lab directly by oscillating from a laser beam, and then be detected by the ``light-shining-through-the-wall'' (LSW) experiments (see Ref.~\cite{Jaeckel:2013ija} for a recent review and references therein). 

Here we review our recent works~\cite{An:2013yfc,An:2013yua} where we have discovered the leading order contribution to the stellar energy loss in the dark photon model, which is from the resonant emission of the longitudinally polarized dark photon. Taking this leading contribution into account the constraints from stellar lifetime is significantly enhanced so that all the current limits of the LSW experiments find their sensitivities deeply inside the exclusion region. We also show the that traditional solar helioscope motivated by axion models are not sensitive to longitudinal mode of dark photon, and the more proper way to detect the longitudinal model of dark photon is to use a detector with a large volume and a large density, which is the dark matter detector. We further show that the XENON10 experiment provides the up-to-date most stringent constraint. 

\section{The dark photon model}

The Lagrangian for a dark $U(1)$ vector field connecting to the SM only through the kinetic mixing with the photon field at energy scale much lower than the weak scale takes the form,
\begin{equation}
  {\cal L} = -\frac{1}{4} F_{\mu\nu}^2-\frac{1}{4} V_{\mu\nu}^2 -
  \frac{\kappa}{2} F_{\mu\nu}V^{\mu\nu}
  + e J_{\mathrm{em}}^{\mu} A_{\mu} ,
\end{equation}
where $J_\mu$ is the electric current and $\kappa$ is the kinetic mixing parameter. For a massive $U(1)$ gauge boson, the mass term can be written as
\begin{equation}
{\cal L}_{\rm mass} = \frac{1}{2} m_V^2 \left(V_\mu - \frac{\partial_\mu a}{m_V}\right)^2 \ ,
\end{equation}
where $a$ is the would-be Goldstone field. There might be an additional light Higgs mode in this theory. In the St\"uckelberg case where there is no such a light Higgs. In the Unitary gauge the mass term can be simplified as
\begin{equation}
  {\cal L}_{\rm mass} = \frac{1}{2} m_V^2 V_\mu V^\mu .
\end{equation}

On the other hand, if there is a light Higgs boson with a mass similar to the dark photon accompanied with the spontaneous symmetry breaking of the dark $U(1)$ field, additional terms of interactions between the Higgs and the dark photon appear, which can be written as
\begin{equation}\label{int}
{\cal L}_{\rm int} = e' m_V h' V_\mu^2 + \frac{1}{2} e'^2 h'^2 V_\mu^2 \ ,
\end{equation}
where $h'$ is the dark Higgs field and $e'$ is the gauge coupling constant of the dark $U(1)$ group. We will see that in the region that $m_V$ is much smaller than the energy scale of the process, the interaction term $e' m_V h' V_\mu^2$ dominates both the production and the detection processes.

\section{St\"uckelberg case}

\subsection{Solar flux}

\begin{figure}
\centering
\includegraphics[height=1.5in]{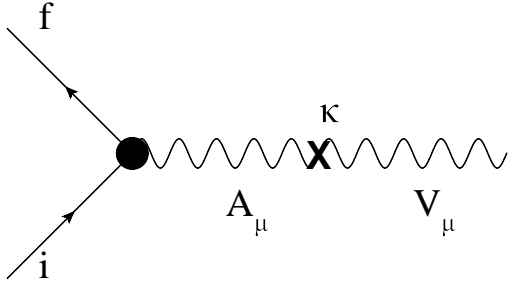}
\caption{Feynman diagram for emission of dark photon in the St\"uckelberg case.}\label{fig:feynman_stu}
\end{figure}

The Feynman diagram for the leading order process for the emission of dark photon is shown in Fig.~\ref{fig:feynman_stu}. To calculate this diagram, we need to remember that this process happens inside Sun, which is a thermal plasma. As a result, the propagation of the photon field acquires a correction from the plasma effect. Using the equation of motion of the dark photon field, the kinetic mixing can be written as $\kappa A_\nu\partial_\mu V^{\mu\nu} = \kappa m_V^2 A_\nu V^\nu$. With the Feynman gauge the propagator of the photon field can be written as 
\begin{equation}
\frac{1}{k^2 - \Pi_{T,L}} = \frac{1}{m_V^2 - \Pi_{T,L}} \ ,
\end{equation}
where $\Pi_{T,L}$ are the projections of the polarization tensor, defined as
\begin{equation}
\Pi^{\mu\nu} = e^2 \langle J^\mu_{\rm em}, J^\nu_{\rm em}\rangle = \Pi_T \epsilon^{T\mu}_i \epsilon^{T\nu}_i + \Pi_L \epsilon^{L\mu} \epsilon^{L\nu} \ .
\end{equation}
where $\epsilon^{T,L}$ are the polarization vectors for the transverse and longitudinal modes of the dark photon respectively. 
Therefore, the matrix element of this process can be written as
\begin{equation}\label{amplitude}
{\cal M} = -\frac{\kappa m_V^2}{m_V^2 - \Pi_{T,L}} \langle f| e J^\mu_{\rm em} |i\rangle \epsilon_\mu^{T,L} \ .
\end{equation}

We know that the longitudinal mode of a massless vector boson cannot couple to a conserved current. Then, it is obvious that for the massive case, we have
$J^\mu_{\rm em} \epsilon_\mu^L \propto m_V$. 
Therefore, we have 
\begin{equation}
\Pi_L \propto  m_V^2\ .
\end{equation}
As a result, for the longitudinal mode, we can see that the $m_V$ dependence of the first factor in Eq.~(\ref{amplitude}) is canceled. Therefore, the $m_V$ scaling of the emission rate of the longitudinally polarized dark photon can be simply written as
\begin{equation}\label{eq8}
\Gamma_L \propto \kappa^2 m_V^2 \ . 
\end{equation}

For transverse modes, however, in the near-vacuum region, we have ${\cal M} \propto m_V^0$; and in the region that $m_V^2 \ll \Pi_T$, we have ${\cal M}\propto m_V^2$. Therefore, the $m_V$ scaling of the emission rate of the transversely polarized dark photon can be simply written as
\begin{equation}\label{eq9}
\Gamma_T \propto \left\{\begin{array}{ll} \kappa^2 m_V^0 \ , & {\rm near~vaccuum} \\ \kappa^2 m_V^4 \ ,&  m_V \ll \omega_p \end{array}\right.
\end{equation}
where $\omega_p$ is the plasma frequency of the media labels the energy scale of the emission process. From Eqs.~(\ref{eq8}) and (\ref{eq9}) we can see that in the small $m_V$ region, the dark emission is dominated by the longitudinal mode. 

Another reason for the longitudinal mode to be the dominant in the dark emission is that it can be effectively produced through the resonance of plasma oscillation. From Eq.~(\ref{amplitude}), we can see that in the case of $m_V^2 = {\rm Re} \Pi_{T,L}$ the transverse (longitudinal) modes is on resonance. Inside a thermal plasma, a resonance means that a thermal bath of photons (transverse and longitudinal) slowly transits into dark photon through the kinetic mixing. To make this transition to happen, the four-momenta must satisfy the on-shell conditions for the photon and the dark photon at the same time. Inside the plasma, the on-shell conditions for the transverse and longitudinal photon field can be written as 
$\omega^2 - |\vec k|^2 = \omega_p^2$ and $\omega^2 = \omega_p^2$, 
respectively. However, for the dark photon field, the on-shell condition is $\omega^2 - |\vec k|^2 = m_V^2$ for both the transverse and longitudinal modes. Then the resonant condition can be written as
\begin{eqnarray}
{\rm Transverse~mode:} ~~~&& m_V = \omega_p \nonumber\\
{\rm Longitudinal~mode:} ~~~&& \omega = \omega_p \nonumber \\
\end{eqnarray}
Inside the Sun, $\omega_p$ varies from 300 eV to 1 eV depending on the radial position. Therefore, the transverse mode can be resonantly produced only when $m_V$ is between 1 eV to 300 eV. Whereas, the longitudinal mode can be resonantly produced as long as $m_V$ is smaller than 300 eV. 
There are also non-resonant production processes, and inside the Sun, the bremsstrahlung process dominates. The detailed discussion can be found in Ref.~\cite{An:2013yfc}. 

\begin{figure}
\centering
\includegraphics[height=2.5in]{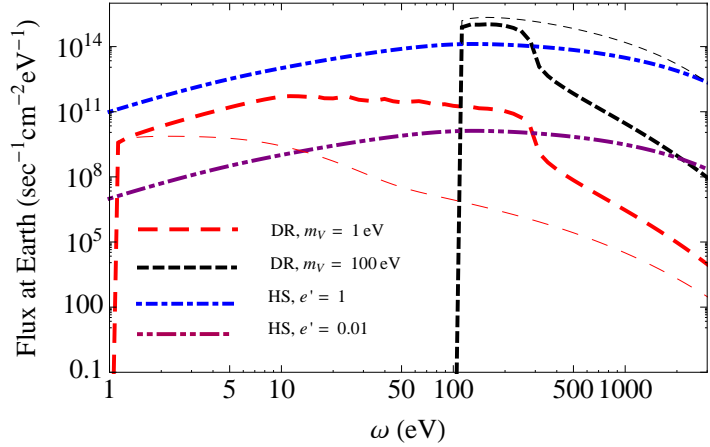}
\caption{Fluxes at the Earth as functions of energy for both
  the SC and HC dark photon for $\kappa = 10^{-12}$. The red and black thick dashed
  curves show the contribution from longitudinal dark radiation (DR) for $m_V=$1 eV
  and 100 eV, respectively. The corresponding thin curves show the transverse contribution. The blue and purple dotted dashed curves show the contribution from the Higgs-strahlung process for $e'=1$ and 0.01, respectively.}\label{fig:flux}
\end{figure}

Using the solar model~\cite{Bahcall:2004pz}, the fluxes of the transverse modes and the longitudinal mode of 1 and 100 eV dark photon emitted from the Sun are shown in Fig.~\ref{fig:flux}. Using the criteria that the energy loss of the Sun from the dark radiation should be smaller than 10\% of its luminosity~\cite{Gondolo:2008dd}, one can get the upper limit on $\kappa$ which is shown in Fig.~\ref{fig:constraints}. One can see that the sensitivities of current LSW experiments and the helioscope experiments are deeply inside the exclusion region. 

From Fig.~\ref{fig:flux}, one can see that the spectrum of the flux is relatively flat in the region where $m_V < 300$ eV, and then drops exponentially. Therefore, the direct detection experiments of solar dark photon should target at this energy region.  
 
\subsection{Direct detection of solar dark photon}

\begin{figure}
\centering
\includegraphics[height=3in]{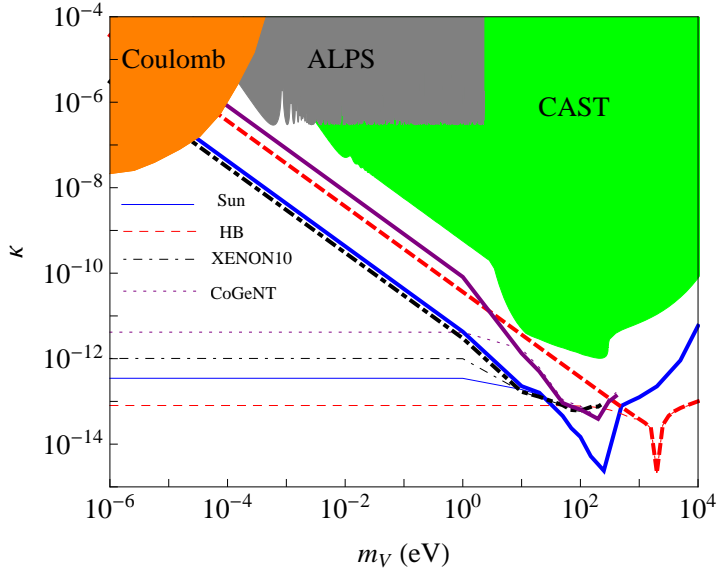}
\caption{Constraints on $\kappa$ as functions of $m_V$. The solid,
  dashed, dot-dashed and dotted curves show constraints from the energy loss of the Sun by requiring that the dark photon luminosty does not exceed 
10\% of the standard solar luminosity~\cite{Gondolo:2008dd}, energy loss of red giant stars (RG), the
  XENON10 experiment and the CoGeNT experiment, respectively. The
  thick curves are for the SC, whereas the thin curves are for the HC
  with $e'=0.1$. For comparison, the current bound (gray shading) 
from the LSW-type experiments are shown (see Ref. \cite{Ehret:2010mh} for
  details). The conservative constraint from the CAST experiment~\cite{Andriamonje:2007ew} by
  considering the contributions from only the transverse
  modes~\cite{Redondo:2008aa} is also shown in green shading.The orange shaded region is excluded from tests of the inverse square law of the Coulomb interaction~\cite{Bartlett:1988yy}.}\label{fig:constraints}
\end{figure}

The expected signal rate for direct detection of solar dark photon can be written as
\begin{equation}
N_{\rm exp} = VT \int_{\omega_{\rm min}}^{\omega_{\rm max}} d\omega \left( \frac{d\Phi_T}{d\omega} \frac{\Gamma_T}{v} + \frac{d\Phi_L}{d\omega} \frac{\Gamma_L}{v} \right) {\rm Br} \ ,
\end{equation}
where $V$ is the volume of the detector, $T$ is the live time of the experiment, $\Phi_{T,L}$ are the fluxes of the transverse and longitudinal modes, $v$ is the velocity of dark photon, and $\rm Br$ is the branching ratio to the desired signal observed in the experiment. 

Fig.~\ref{fig:feynman_stu} also shows the leading contribution of the Feynman diagram for the absorption of dark photon, and we can also use Eq.~(\ref{amplitude}) to calculate the matrix element by replacing $\omega_p^2$ with $\omega^2\Delta\varepsilon_r$, where $\Delta\varepsilon_r = \varepsilon_r - 1$ and $\varepsilon_r$ is the relative permittivity of the material. The total absorption rate can be written as
\begin{eqnarray}
\Gamma_T &=& { \left( \frac{\kappa^2m_V^4{{\rm Im} \varepsilon_r}}{\omega^3|\Delta\varepsilon_r|^2} \right)  }
{\left[1 + \frac{2 m_V^2 \omega^2 {\rm Re}\Delta\varepsilon_r + m_V^4}{\omega^4 |\Delta\varepsilon_r|^2}\right]^{-1}} \!\!\!\!\! , 
\nonumber\\
\Gamma_L &=& \frac{\kappa^2 m_V^2 {\rm Im} \varepsilon_r}{\omega|\varepsilon_r|^2} \ .
\label{gamma2}\end{eqnarray}
In the region that $m_V^2 \gg \omega^2 {\rm Re}\Delta\epsilon_r$, $\Gamma_T$ can be simplified as $\Gamma_T = \kappa^2 \omega {\rm Im}\varepsilon_r$, and in the region that $m_V^2 \ll \omega^2 {\rm Re}\Delta\epsilon_r$, we have $\Gamma_T = \kappa^2 \omega\left(\frac{m_V^2}{\omega^2|\Delta\varepsilon_r|}\right)^2 {\rm Im}\varepsilon_r$. We know that both the real and imaginary parts of $\Delta\varepsilon_r$ are proportional to the atomic number density of the material $n_A$. Therefore $\Gamma_T$ is proportional to $n_A$ in the large $m_V$ region and proportional {\it inversely} to $n_A$ in the small $m_V$ region, whereas$\Gamma_L$ is always proportional to $n_A$. However, from the analysis of the last subsection we know that in the small $m_V$ region, the dark flux is dominated by the longitudinal mode. Therefore, the signal rate of the solar dark photon is always proportional to $n_A$. As a consequence, the detectors of solar dark photon should be built with large density materials. For similar reasons dark matter detectors are made from such materials, and the S2 only analysis of the XENON10 experiment~\cite{Angle:2011th} is just in the desired energy region as the solar dark photon. A detailed analysis can be found in Ref.~\cite{An:2013yua}, and the constraint is shown in Fig.~\ref{fig:constraints}, where the stellar constraints are also shown. One can see that for the St\"uckelberg case, the sensitivity of XENON10 S2 only analysis has already surpassed the constraint from the solar lifetime. Further dark matter experiments may have a chance to detect the dark photon from the Sun.

\section{The Higgsed case}

\begin{figure}
\centering
\includegraphics[height=1.5in]{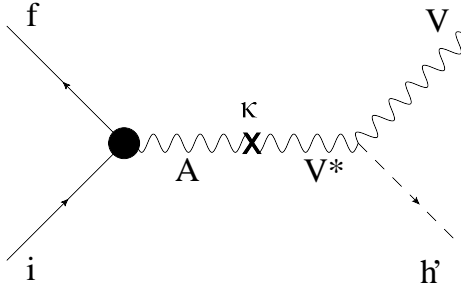}
\caption{Feynman diagram for emission of dark photon in the Higgsed case.}\label{fig:feynman_higgs}
\end{figure}

\subsection{Solar flux of dark photon}

In the Higgsed case, due to the interaction terms (\ref{int}), there are additional diagrams contributed to the emission of the dark photon from the Sun which is the Higgsstrahlung process shown in Fig.~\ref{fig:feynman_higgs}. This process is suppressed by the phase space of the two-body final state. However, in the small $m_V$ region, the one-body final state process shown in Fig.~\ref{fig:feynman_stu} suffers from suppression of $m_V^2/\omega_p^2$ and $m_V^4/\omega^4$ for the longitudinal and transverse modes respectively. For the Higgsstrahlung process, this suppression does not appear. To see this, let's consider the limit that $m_V \ll \omega$. In this region, we can use the Goldstone equivalence theorem to estimate the Higgsstrahlung process. Furthermore, it is easy to see that the photon propagator can always be on-shell. Therefore inside the plasma, the Higgsstrahlung process can be seen as the process that a {\it massive} photon decays into a pair of massless dark Higgs particles through the kinetic mixing with the dark photon field. Take the transverse photon as an example, in this resonant case, since the {\it massive} photon is on shell, the kinetic mixing can be written as $\kappa V_\nu \partial_\mu F^{\mu\nu} = \kappa \omega_p^2 A_\mu V^\mu$. And since $m_V \ll \omega_p$, the propagator of $V$ can be estimated as $1/\omega_p^2$. Therefore, this process is equivalent to the process of a mass vector boson with mass $\omega_p$ decays into a pair of scale particle with an {\it effective} charge of $\kappa e'$. As a result, this process does not suffer from the suppression from $m_V$. Therefore, in the region that $m_V$ is comparable $\omega_p$, the Higgsstrahlung process is sub-dominant due to the phase-space suppression. Whereas it is dominant in the region that $m_V \ll \omega_p$ due to the lack of the $m_V$ suppression. The flux of the dark photon in the Higgsed case is shown in Fig.~\ref{fig:flux}, where one can see that the maximum of the flux is around 300 eV. Therefore, the S2 only analysis of the XENON10 result is also sensitive to the Higgsed case. 

\subsection{Direct detection of solar dark photon}
 
\begin{figure}
\centering
\includegraphics[height=1.5in]{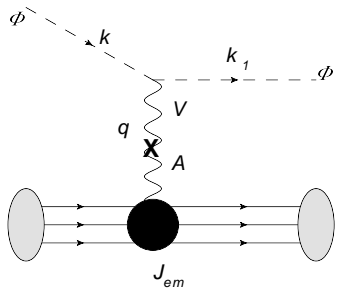}
\caption{Feynman diagram for emission of dark photon in the Higgsed case.}\label{fig:higgsdetect}
\end{figure} 
 
The interaction terms (\ref{int}) also contributes to the direct detection of the dark photon. In the region where $m_V\ll \omega$, using the Goldstone equivalent theorem the Feynman diagram for this additional direct detection process is shown in Fig.~\ref{fig:higgsdetect}, where $\Phi$ is the Higgs field before the spontaneous breaking of the dark $U(1)$ gauge symmetry. From the Feynman diagram, one can see that this scatter process is very similar to the diagram of the deep-inelastic scattering, where all the detailed information of the nuclear substructure are parameterized in the parton distribution function. Similarly, in this process, the information of the atomic structures are parameterized by relativepermittivity of the material. Using the dispersion relation one can show~\cite{An:2013yua} that the differential scattering rate with the respect to the energy transfer to atoms $q^0$ at the limit that $m_V \rightarrow 0$ is given by 
\begin{equation}
\label{GHC}
\frac{d\Gamma}{d q^0} \approx \frac{ \kappa^2 e'^2 }{4\pi^2} \frac{k_1^0 - q^0}{k_1^0} 
\left[ \log\left( \frac{4k_1^0(k_1^0 - q^0)}{{(q^0)}^2 |\Delta\varepsilon_r|} \right) - 1\right] {\rm Im}\varepsilon_r(q^0) ,
\end{equation}
where $k_1^0$ is the energy of the incoming dark photon. The collinear divergence in this case is regularized by the {\it effective mass} of the photon inside the material. Combined the contributions from both this scattering process and the absorption process discussed in the St\"uckelberg case, the upper limits on $\kappa$ from dark matter direct detection experiments for the Higgsed case are shown in Fig.~\ref{fig:constraints}, where we can see that in this case, in the very small $m_V$ region, the upper limit on $\kappa$ is independent of the $m_V$. 
In this case, the XENON10 S2 only analysis is less sensitive than the constraints from the lifetime of the horizontal branch stars.

\Acknowledgements

H.A. would like to thank the organizers for organizing this conference, and the hospitality of University of Hawaii.  This research was
supported in part by Perimeter Institute for Theoretical
Physics. Research at Perimeter Institute is supported
by the Government of Canada through Industry Canada
and by the Province of Ontario through the Ministry of
Research and Innovation.

\end{document}